\documentclass[twocolumn,showpacs,floatfix,superscriptaddress,amsmath,amssymb]{revtex4}
\usepackage{color}
\usepackage{amsmath}
\usepackage{amsfonts}
\usepackage{bbm}
\usepackage{mathrsfs}
\usepackage{txfonts}
\usepackage{amssymb}
\usepackage{graphicx}
\usepackage{hyperref}
\usepackage{ulem}
 \usepackage{overpic}
 \usepackage{psfrag}
\usepackage{tabularx}
\usepackage{array}
\usepackage{placeins}
\newcommand{\be}{\begin{equation}}
\newcommand{\ee}{\end{equation}}
\newcommand{\PreserveBackslash}[1]{\let\temp=\\#1\let\\=\temp}

\begin{document}
\title{Quantum fidelity approach to detecting quantum phases: revisiting the bond alternating Ising chain}

\author{Hai Tao Wang}
\affiliation{Centre for Modern Physics and Department of Physics,
Chongqing University, Chongqing 400044, The People's Republic of
China}

 \author{Sam Young Cho}
 \email{sycho@cqu.edu.cn}
\affiliation{Centre for Modern Physics and Department of Physics,
Chongqing University, Chongqing 400044, The People's Republic of
China}

 \author{Murray T. Batchelor}
\affiliation{Centre for Modern Physics and Department of Physics,
Chongqing University, Chongqing 400044, The People's Republic of
China}
\affiliation{Mathematical Sciences Institute and Department of Theoretical Physics,
Research School of Physics and Engineering, Australian National University, Canberra ACT 0200, Australia}

\begin{abstract}
We demonstrate the quantum fidelity approach
for exploring and mapping out quantum phases.
As a simple model exhibiting a number of distinct quantum phases, we consider the
 alternating-bond Ising chain using the infinite time evolving block decimation method
 in the infinite matrix product state representation.
Examining the quantum fidelity with an arbitrary reference state in
the whole range of the interaction parameters leads to the explicit detection of the 
doubly degenerate groundstates, indicating a $Z_2$ broken symmetry.
 The discontinuities of the fidelity indicate a first-order quantum phase transition
 between the four ordered phases.
 In order to characterize each phase, based on the spin configurations from the spin correlations,
 even and odd antiferromagnetic order parameters are introduced.
 The four defined local order parameters are shown to characterize each phase
 and to exhibit first-order quantum phase transitions between the ordered phases.
\end{abstract}

\pacs{03.65.Vf, 64.70.Tg, 75.10.Pq, 75.30.Kz,75.40.Mg}

\maketitle

\section{Introduction}

 Tensor network representations
 have enabled significant progress in the computational study of quantum phase transitions
  \cite{Fannes,Ostlund,Perez,White,Schollwock,Vidal03,Vidal07,Nurg,Zhou4,Orus}.
 More specifically, a wave function represented by a tensor network is convenient for the
 simulation of quantum many-body systems.
 In one-dimensional spin systems,
 a wave function for infinite-size lattices can be described by
 the infinite matrix product state (iMPS) representation \cite{Vidal03,Vidal07}.
 The iMPS and tensor networks in general offer an understanding of critical
 phenomena in infinite and finite lattice systems from the
 perspective of quantum entanglement (see, e.g., Refs~\cite{Schollwock,Perez,Orus} for reviews).
 It has also been demonstrated that quantum fidelity is a useful tool to detect
 phase transition points and
 degenerate groundstates \cite{Zhou1,Zhao,Su} originating from a spontaneous symmetry breaking
 for a broken symmetry phase, without knowing what type of internal order is
 present in quantum many-body states.
 Quantum fidelity based approaches have been successfully implemented
 to investigate quantum phase transitions in a number of models.
 Examples in one-dimension include the Ising model in a transverse magnetic field \cite{Zhou2}, 
 %and with antisymmetric, anisotropic and alternating bond interactions \cite{Li},
 the XYX model in an external magnetic field \cite{Zhao},
 the bond alternating spin-1/2 Heisenberg chain \cite{Wang}, 
 and the $q$-state Potts quantum chain \cite{Su}.

In this study we further demonstrate the quantum fidelity approach
for exploring and mapping out quantum phases.
As a simple but illustrative model exhibiting a number of distinct quantum phases, we consider the
Ising chain with alternating interaction strengths.
This bond alternating model, also known as the dimerized Ising chain, has been used, for example, to
 investigate non-equilibrium spin dynamics at finite temperatures~\cite{Cornell1,Cornell2},
 the so called Glauber dynamics \cite{Glauber}.
More recently it has also been investigated in the context of
additional Dzyaloshinskii-Moriya interactions \cite{DM1,DM2,DM3}.
However, it has not been fully considered how Landau's spontaneous symmetry
breaking picture applies to this model.
Here we address this issue using the quantum fidelity approach and demonstrate how to define
the explicit order parameters quantifying the distinct quantum phases throughout the whole parameter range.

 We calculate the groundstate wavefunction of the infinite spin-$1/2$ bond-alternating Ising chain
 by employing the infinite matrix product state (iMPS) representation~\cite{Vidal03,Vidal07}
 with the infinite time evolving block decimation (iTEBD) method developed by Vidal~\cite{Vidal07}.
 In order to capture the symmetry-broken phases, we use period four
 matrix product states including characteristic eight tensors
 for the iMPS representation.
 The basic idea outlined in Sec.~II is to use the quantum fidelity with an arbitrary reference state,
allowing the doubly degenerate groundstates to be detected for the whole range of
 the two exchange interaction parameters.
 From the discontinuities of the quantum fidelity,
 we find that a first-order phase transition occurs between the ordered phases
 once one of the two interaction strengths changes sign.
 Further, from the spin correlation functions calculated from the degenerate ground states,
 we discuss in Sec.~III the characteristic spin configuration for each phase
 and define the possible local order parameters in Sec.~IV, including
 even and odd antiferromagnetic ordering.
 It is shown that the four defined local order parameters
 reveal four ordered phases and exhibit first-order transitions between
 the ordered phases in agreement with the results using the quantum fidelity.
Concluding remarks are given in Sec.~V.

\section{Bond-alternating Ising chain and quantum fidelity per site}

 We consider the  spin-$1/2$ bond-alternating Ising chain given by the Hamiltonian
\begin{equation}
H= \sum_{i=-\infty}^\infty \left(J' \, S^{z}_{2i-1}S^{z}_{2i}+ J \, S^{z}_{2i}S^{z}_{2i+1}\right),
\label{Hamt}
\end{equation}
 where $S^z_i$ is the spin operator on the $i$th site.
 The exchange couplings are $J$ and $J'$, which we parametrise in terms of the variable $\theta$,
 with $J=\cos\theta$ and $J'=\sin\theta$.
 For $J=J' < 0$ ($\theta=5\pi/4$) the system becomes the ferromagnetic (FM) Ising model
 and for $J=J'>0$ ($\theta=\pi/4$) the antiferromagnetic (AFM) Ising model.
 In the both cases $J=J' > 0$ and $J=J' < 0$ the Hamiltonian is
 one-site translational invariant.
 Due to the bond alternation, then, the Hamiltonian is two-site translational invariant, except for
 the cases $J=J' > 0$ and $J=J' < 0$.

%
%%%%%%%%%%%%%%%%%%%%%%%%%%%fig 1%%%%%%%%%%%%%%%%%%%%%%%%%%%%%%%%%%%%
\begin{figure}
\includegraphics [width=0.45\textwidth]{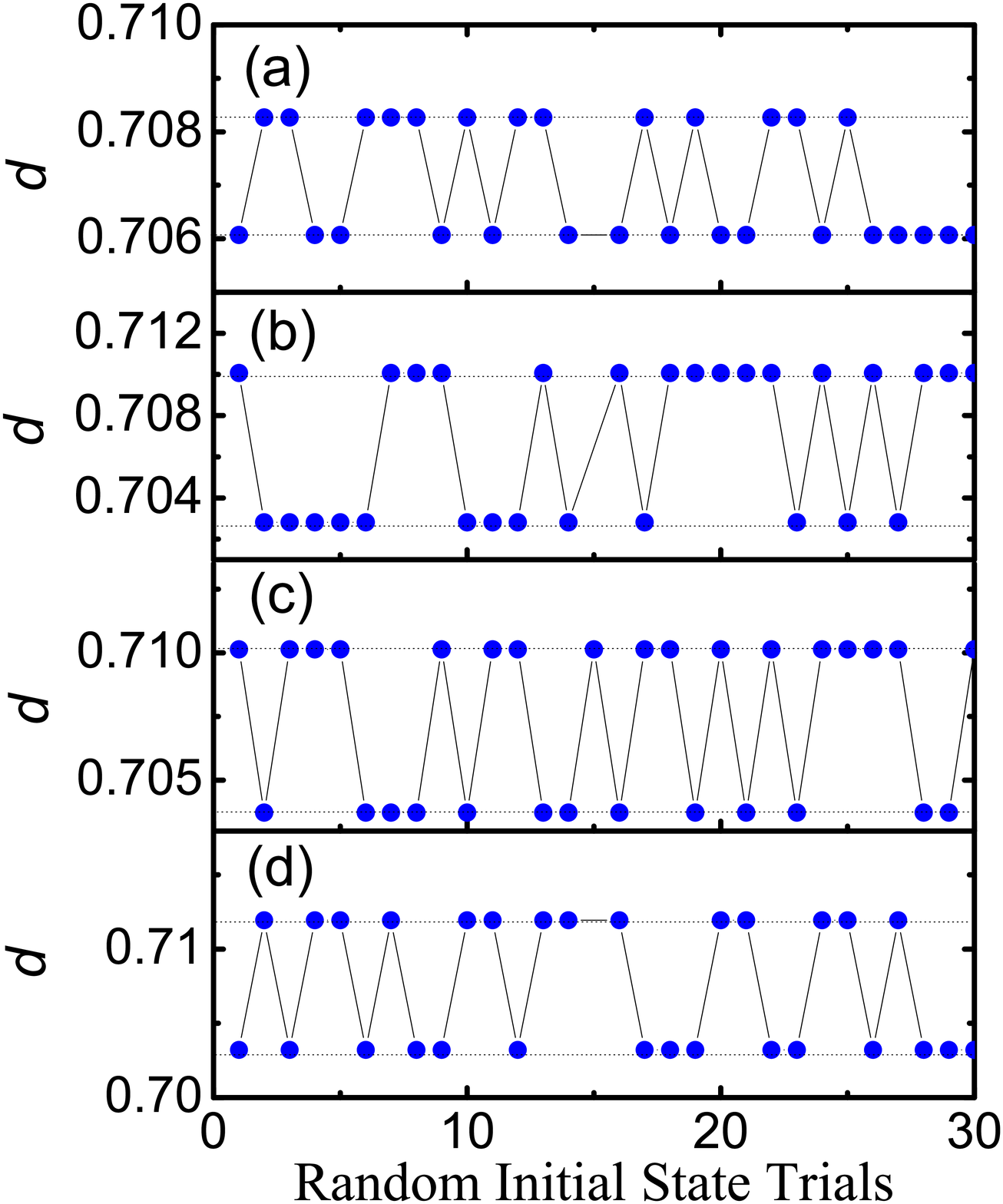}
\caption{ (Color online)
 Fidelity per site $d(\Psi^{(n)}(\theta),|\phi\rangle)$ between the $n$-th calculated groundstate $|\Psi^{(n)}(\theta)\rangle$
 from the $n$-th random initial state and
 an arbitrary chosen reference state $|\phi\rangle$ for
 (a) $\theta=\pi/4$, (b) $\theta=3\pi/4$,
 (c) $\theta=5\pi/4$ and (d) $\theta=7\pi/4$.
The horizontal axis denotes the number $n$ of the random initial
state.
 It is clearly shown that for each case there exist two degenerate groundstates corresponding to
 the two values of $d$.
}
 \label{fig1}
\end{figure}
%%%%%%%%%%%%%%%%%%%%%%%%%%%fig 1%%%%%%%%%%%%%%%%%%%%%%%%%%%%%%%%%%%%

An iMPS ground state of the system can be obtained by
using the iTEBD algorithm with a chosen initial state in the iMPS representation.
We calculate $n$ groundstate values $|\Psi^{(n)}(\theta)\rangle$ corresponding to the $n$-th random initial state.
 In order to determine how many groundstates exists for a given parameter $\theta$,
 we consider the quantum fidelity
 $F(|\Psi^{(n)}(\theta)\rangle,|\phi\rangle)=|\langle\Psi^{(n)}(\theta)|\phi\rangle|$
which is the overlap function between the $n$-th calculated ground state $|\Psi^{(n)}(\theta)\rangle$
 and an arbitrary reference state $|\phi\rangle$.
 For our numerical study, the reference state $|\phi\rangle$ is chosen randomly.
 The quantum fidelity scales as $F(|\Psi^{(n)}(\theta)\rangle,|\phi\rangle)\sim d^L$,
 where $L$ is the system size.
 Following, e.g., Ref.~\onlinecite{Su}, the fidelity per site can be defined as
\begin{equation}
 \ln d(|\Psi^{(n)}(\theta)\rangle,|\phi\rangle)
 = \lim_{L \rightarrow \infty} \frac {\ln
  F(|\Psi^{(n)}(\theta)\rangle,|\phi\rangle)}{L}.
 \label{FLS}
\end{equation}
 From the fidelity  $F(\Psi^{(n)}(\theta)\rangle,|\phi\rangle)$, the fidelity per site satisfies
 the properties: (i) normalization $d(|\phi\rangle,|\phi\rangle)=1$
 and (ii) range $0 \leq d(\Psi^{(n)}(\theta)\rangle,|\phi\rangle) \leq 1$.

A degenerate groundstate can be determined from the fidelity per site $d(|\Psi^{(n)}\rangle,|\phi\rangle)$
as a function of the random initial state trials $n$,
as shown in Fig.~\ref{fig1} for (a) $\theta=\pi/4$, (b) $\theta=3\pi/4$,
 (c) $\theta=5\pi/4$, and (d) $\theta=7\pi/4$.
 For the iMPS representation, the truncation dimension $\chi$ used is $\chi=32$.
These plots show that there are two different values of the fidelity per site for the groundstates
from the $30$ random initial states.
For a large enough number of random initial state trials, the probability $P(n)$ of each
 degenerate groundstate approaches $1/2$, i.e., $\lim_{n \rightarrow \infty} P(n)=1/2$.
This implies that there are doubly degenerate groundstates for each given $\theta$ value.

 %%%%%%%%%%%%%%%%%%%%%%%%%%%fig 2%%%%%%%%%%%%%%%%%%%%%%%%%%%%%%%%%%%%
\begin{figure}
\includegraphics [width=0.45\textwidth]{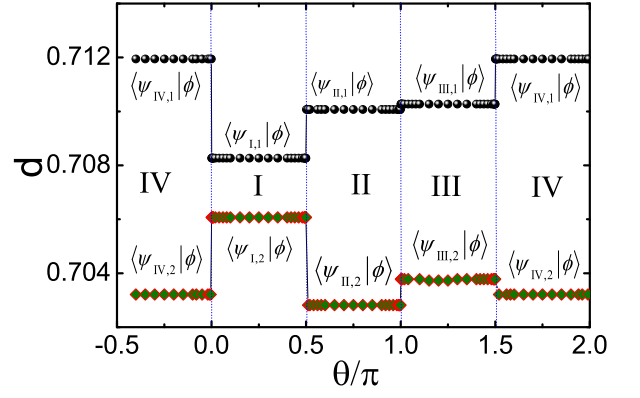}
\caption{ (Color online)
 Fidelity per site between the doubly degenerate
 groundstates with the same reference state used in Fig. \ref{fig1} as a function of the interaction parameter $\theta$.
 The four different ordered phases $\{\mathrm{I,II,III,IV}\}$ are clearly distinguished.
 For each ordered phase, the two values of the fidelity per site indicate a doubly degenerate groundstate.
}
 \label{fig2}
\end{figure}
%%%%%%%%%%%%%%%%%%%%%%%%%%%fig 2%%%%%%%%%%%%%%%%%%%%%%%%%%%%%%%%%%%%

 In order to determine how many ordered phases there are in the model,
 we calculated the fidelity per site as a function of the interaction parameter $0\leq \theta \leq 2\pi$ (see Fig.~\ref{fig2}).
 Here, we have chosen the same reference state $|\phi\rangle$ as in Fig.~\ref{fig1}.
 Fig.~\ref{fig2} shows clearly that the fidelity is discontinuous at four points, i.e., $\theta=0$, $\pi/2$, $\pi$, and $3\pi/2$.
 The discontinuous fidelities indicate that a first-order quantum phase transition occurs
 at each discontinuous point corresponding to each critical point.
 Consequently, the system has four ordered phases:

 \begin{itemize}
 \item[] I: \quad $0 < \theta < \pi/2$ $(J > 0 \mbox{~and~} J' > 0)$,
  \item[] II: \quad $\pi/2 < \theta < \pi$ $(J < 0 \mbox{~and~} J' > 0)$,
  \item[] III: \quad $\pi < \theta < 3\pi/2$ $(J < 0 \mbox{~and~} J' < 0)$,
  \item[] IV: \quad $3\pi/2 < \theta < 2\pi$ $(J > 0 \mbox{~and~} J' < 0)$.
  \end{itemize}

 Each phase has doubly degenerate groundstates denoted by
 $|\Psi_{\alpha,1}(\theta)\rangle$ and $|\Psi_{\alpha,2}((\theta)\rangle$,
 where $\alpha \in \{\mathrm{ I,II,III,IV}\}$ labels the phases.
 According to the spontaneous symmetry breaking theory,
 the doubly degenerate groundstates imply
 that a $Z_2$ symmetry is broken for each phase.
 Hence, for each phase, a different $Z_2$ symmetry is broken
 and the system state belongs to a different ordered phase.

\section{Spin correlations and groundstate wavefunctions}

%%%%%%%%%%%%%%%%%%%%%%%%%%%fig 3%%%%%%%%%%%%%%%%%%%%%%%%%%%%%%%%%%%%
\begin{figure}
\includegraphics [width=0.4\textwidth]{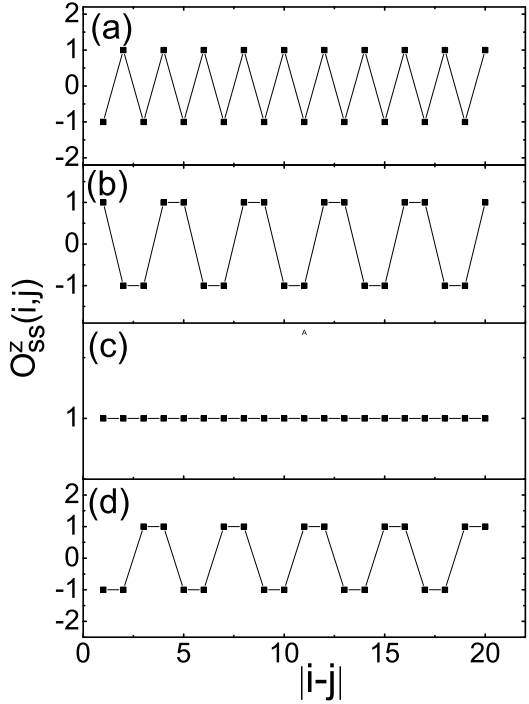}
\caption{
 The spin correlation $O^{z}_{ss}(i,j)$ as a function of the site distance $|i-j|$
 for (a) $\theta=\pi/4$, (b) $\theta=3\pi/4$, (c) $\theta=5\pi/4$, (d) $\theta=7\pi/4$.}
 \label{fig3}
\end{figure}
%%%%%%%%%%%%%%%%%%%%%%%%%%%fig 3%%%%%%%%%%%%%%%%%%%%%%%%%%%%%%%%%%%%

 In order to gain further insight into the four ordered phases,
 we consider the spin-spin correlations for each phase.
 The spin-spin correlation function is defined by
\begin{equation}
 O_{ss}^{z}(i,j)=\langle \sigma^{z}_i \, \sigma^{z}_j \rangle,
 \label{SS}
\end{equation}
 where $i$ and $j$ denote locations along the chain.
 In Fig.~\ref{fig3}, we plot the spin-spin correlation $O_{ss}^{z}(i,j)$
 as a function of the separation distance $|i-j|$ for the typical parameter values
 (a) $\theta=\pi/4$, (b) $\theta=3\pi/4$, (c) $\theta=5\pi/4$, and (d) $\theta=7\pi/4$ in each of the phases.

 \subsection{Antiferromagnetic phase}
 
The two groundstates for each phase give the same spin-spin correlation.
 Furthermore, the spin-spin correlation as a function of the lattice distance
 has only two values, i.e., $+1$ or $-1$.
 This result implies that the direction of the two spins $i$ and $j$ is either parallel for
 $O_{ss}^{z}(|i-j|)= +1$ or anti-parallel for $O_{ss}^{z}(|i-j|)= -1$.
 For instance, for $\theta=\pi/4$, the spin configurations, using an obvious notation,
are either
  $\cdots \uparrow_{i} \downarrow_{i+1} \uparrow_{i+2} \downarrow_{i+3} \uparrow_{i+4}\cdots$ or
  $\cdots \downarrow_{i} \uparrow_{i+1} \downarrow_{i+2} \uparrow_{i+3} \downarrow_{i+4}\cdots$.
 Consequently, for $\theta=\pi/4$,
 the groundstate wavefunctions can be written as
\begin{subequations}
\begin{eqnarray}
 |\Psi_{\mathrm{I},1 }(\pi/4)\rangle
  &=& \prod_{i=-\infty}^{\infty}
            \left|\uparrow\right\rangle_{2i}\, \left| \downarrow\right\rangle_{2i+1},
\\
 |\Psi_{\mathrm{I},2} (\pi/4)\rangle
  &=& \prod_{i=-\infty}^{\infty}
            \left|\downarrow\right\rangle_{2i}\, \left| \uparrow\right\rangle_{2i+1}.
 \label{AF}
\end{eqnarray}
\end{subequations}
  Thus, for $0 < \theta < \pi/2$ ($J > 0$ and $J'>0$),
 the groundstate is in the antiferromagnetic (AFM) phase.
 The groundstate wavefunctions are two-site translational invariant.
 However, the Hamiltonian in Eq. (\ref{Hamt}) is two-site translational invariant
 for $J \neq J'$ but is one-site translation invariant for $J = J'$.
 Also, note that one of the two groundstates transforms to the other groundstate
 under the spin-flip transformation.
  Hence, for $J \neq J'$, the two groundstates result from the spontaneous symmetry breaking
 of the spin-flip symmetry.
We can think of the two groundstates in the AFM phase as losing more symmetry for $J=J'$ than 
$J \neq J'$ from the spontaneous symmetry breaking of the spin-flip symmetry.

 \subsection{Odd antiferromagnetic phase}
 
 For $\theta=3\pi/4$, the spin correlation in Fig. \ref{fig3}(b) implies that the spin configuration is either
  $\cdots \uparrow_{i} \uparrow_{i+1} \downarrow_{i+2} \downarrow_{i+3} \uparrow_{i+4} \cdots$ or
 $\cdots \downarrow_{i} \downarrow_{i+1} \uparrow_{i+2} \uparrow_{i+3} \downarrow_{i+4} \cdots$.
 The groundstate wavefunctions are
 \begin{subequations}
\begin{eqnarray}
 |\Psi_{\mathrm{II},1 }(3\pi/4)\rangle
  &=& \prod_{i=-\infty}^{\infty}
            \left|\uparrow\right\rangle_{4i}\, \left| \uparrow\right\rangle_{4i+1}
            \left|\downarrow\right\rangle_{4i+2}\, \left|\downarrow\right\rangle_{4i+3},
\\
 |\Psi_{\mathrm{II},2} (3\pi/4)\rangle
  &=& \prod_{i=-\infty}^{\infty}
            \left|\downarrow\right\rangle_{4i}\, \left|\downarrow\right\rangle_{4i+1}
            \left|\uparrow\right\rangle_{4i+2}\, \left|\uparrow\right\rangle_{4i+3}.
\end{eqnarray}
\end{subequations}
 Thus, for $\pi/2 < \theta < \pi$ ($J < 0 \mbox{~and~} J'>0$),
 the groundstate wavefunctions are four-site translational invariant because
 the exchange interaction on the even bonds is ferromagnetic (FM), i.e., $J < 0$,
 while the exchange interaction on the odd bonds is AFM, i.e., $J' > 0$.
 Note that one of the two groundstates transforms to the other
 groundstate under the spin-flip transformation and the two-site translational transformation.
 The Hamiltonian is invariant for the spin-flip transformation
 and the two-site translational transformation.
 Hence,
 the two degenerate ground states arise from the spontaneous symmetry breaking of the group
 $G = Z_2 \times Z_2 /Z_2$, i.e., the spin-flip $Z_2$ and
 the two-site translational $Z_2$ symmetries.
 The spin configuration shows two-spin alternating behavior.
 In order to distinguish this phase from the AFM phase in region I,
 we then call this phase the odd AFM phase.

\subsection{Ferromagnetic phase}

 The spin correlation for $\theta = 5\pi/4$ in Fig.~\ref{fig3}(c) is FM.
For $\pi < \theta < 3\pi/2$ ($J < 0$ and $J'<0$)
the groundstate wavefunctions are the FM wavefunctions
\begin{subequations}
\begin{eqnarray}
%
%%%%%%%%%%%%%%%%%%%%%%%%%%%%%%%%%%%%%
%%%%%%%%%%%%%%%%%%%%%%%%%%%%%%%%%%%%%
 |\Psi_{\mathrm{III},1 }(5\pi/4)\rangle
  &=& \prod_{i=-\infty}^{\infty}
            \left|\uparrow\right\rangle_{i},
\\
 |\Psi_{\mathrm{III},2} (5\pi/4)\rangle
  &=& \prod_{i=-\infty}^{\infty}
            \left|\downarrow\right\rangle_{i}.
\end{eqnarray}
\end{subequations}
 The groundstate wavefunctions are one-site translational invariant.
 However, the Hamiltonian in Eq.~(\ref{Hamt}) is two-site translational invariant
 for $J \neq J'$ but is one-site translation invariant for $J = J'$.
 Also, note that one of the two groundstates transforms to the other groundstate
 under the spin-flip transformation.
 Hence, for $J \neq J'$, the two groundstates result from the spontaneous symmetry breaking
 of the spin-flip symmetry.
 For $J=J'$, interestingly, the two groundstates in the FM phase have more symmetry
 than the Hamiltonian because the two groundstates have one-site translational symmetry
 but the Hamiltonian has two-site translational symmetry.
 Such a symmetry, which is not directly manifest in the Hamiltonian, is an example of an enhanced or {emergent symmetry}
 \cite{emergent,emergent2,Schmaliah,Batista,Liu,Silvi,Chen}.

 \subsection{Even antiferromagnetic phase}

The spin correlation for $\theta = 7\pi/4$ in Fig. \ref{fig3}(d) is similar to the case $\theta = 3\pi/4$.
Here the groundstate wavefunctions are
 \begin{subequations}
\begin{eqnarray}
 |\Psi_{\mathrm{IV},1 }(7\pi/4)\rangle
  &=& \prod_{i=-\infty}^{\infty}
            \left|\uparrow\right\rangle_{4i}\, \left| \downarrow\right\rangle_{4i+1},
            \left|\downarrow\right\rangle_{4i+2}\, \left|\uparrow\right\rangle_{4i+3},
\\
 |\Psi_{\mathrm{IV},2} (7\pi/4)\rangle
  &=& \prod_{i=-\infty}^{\infty}
            \left|\downarrow\right\rangle_{4i}\, \left|\uparrow\right\rangle_{4i+1},
            \left|\uparrow\right\rangle_{4i+2}\, \left|\downarrow\right\rangle_{4i+3}.
\end{eqnarray}
\end{subequations}
 Thus for $3\pi/2 < \theta < 2\pi$ ($J > 0$ and  $J'<0$)
 the exchange interaction on the odd bonds is FM ($J'< 0$)
 while the exchange interaction on the even bonds is AFM ($J > 0$).
 % %
 The groundstate wavefunctions are four-site translational invariant.
 Note that one of the two groundstates transforms to the other
 groundstate under the spin-flip transformation and the two-site translational transformation.
 Also, the Hamiltonian is invariant for the spin-flip transformation
 and the two-site translational transformation.
 Hence,
 the two degenerate ground states arise from the spontaneous symmetry breaking of the group
 $G = Z_2 \times Z_2 /Z_2$, i.e., the spin-flip $Z_2$ and
 the two-site translational $Z_2$ symmetries.
 The spin configuration shows two-spin alternating behavior.
 In order to distinguish this phase from the AFM phase in the regions I and II,
 we then call this phase the even AFM phase.
 However, both of the odd and even AFM phases
 originate from the breaking of the same symmetry.
 Then one needs to consider how to distinguish the two phases.
 This point is clarified in the discussion of the appropriate order parameters in Sec. IV.

%%%%%%%%%%%%%%%%%%%%%%%%%%%fig 4%%%%%%%%%%%%%%%%%%%%%%%%%%%%%%%%%%%%
\begin{figure}
\includegraphics [width=0.4\textwidth]{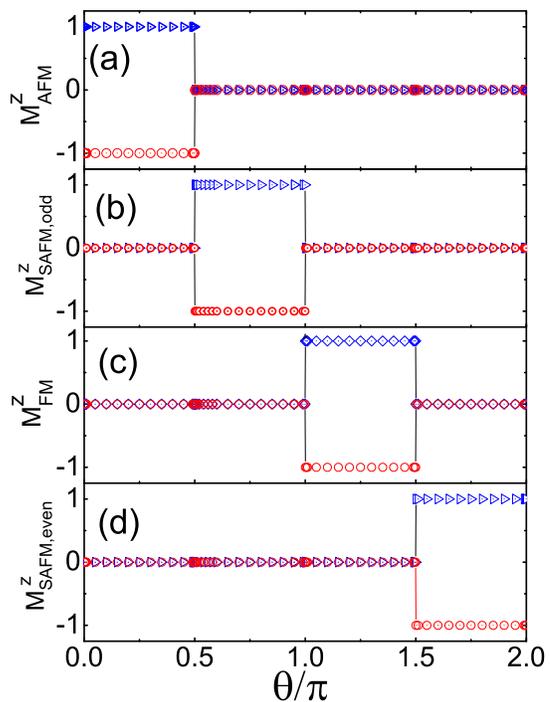}
\caption{ (Color online) The four order parameters as a function of
the interaction parameter $\theta$.
 In (a)-(d), the blue triangles and the red circles denote
 the average value of the local order parameters, defined in the text,
 with each ground state, $|\Psi_{\alpha,1}\rangle$ and $|\Psi_{\alpha,2}\rangle$, respectively.
}
 \label{fig4}
\end{figure}
%%%%%%%%%%%%%%%%%%%%%%%%%%%fig 4%%%%%%%%%%%%%%%%%%%%%%%%%%%%%%%%%%%%

\section{Order parameters}

 To distinguish the four phases, one has to define characteristic local order parameters.
 Based on the spin configurations or the wavefunctions,
 we define the order parameters for the four phases.
 Normally, one uses the FM and the AFM order parameters defined by
 the magnetization $M^{F}_i=\langle \sigma^z_i+\sigma^z_{i+1}\rangle/2$ and
 the staggered magnetization $M^{AF}_i=\langle \sigma^z_i-\sigma^z_{i+1} \rangle/2$.
 However, for the bond-alternating model, these definitions for the order parameters
 do not distinguish between all four phases,
 because of the even AFM and odd AFM phases.
 Furthermore, the odd and even AFM phase
 originate from the spontaneous breaking of the same symmetries.
 To overcome this, one needs to consider the symmetries of the
 groundstates.
 One can notice that from the two degenerate groundstates for each phase,
 the defined FM order parameter should be one-site translational invariant,
 the AFM order parameter two-site translational invariant,
 and both the odd and even AFM order parameters four-site
 translational invariant.
 Then, proper order parameters should be four-site
 translational invariant.
 From the properties of the groundstates for each phase,
 in terms of the normal definitions of the magnetization and the
 staggered magnetization,
 we define four order parameters as follows,
\begin{subequations}
 \begin{eqnarray}
 M^z_{AFM}   &=&   \frac{1}{2}
   \left( M^{AF}_{2i} +M^{AF}_{2i+2} \right) , \label{AFM}   \\
 M^z_{FM}  &=&
   \frac{1}{2}
   \left( M^{F}_{2i}+M^{F}_{2i+2} \right) , \label{FM} \\
 M^z_{AFM,even}   &=&   \frac{1}{2}
    \left( M^{F}_{2i}-M^{F}_{2i+2} \right)  , \label{ESAFM} \\
 M^z_{AFM,odd}   &=&
   \frac{1}{2}
   \left( M^{F}_{2i+1}-M^{F}_{2i+3} \right). \label{OSAFM}
\end{eqnarray}
\end{subequations}

\begin{figure}[t]
\includegraphics [width=0.25\textwidth]{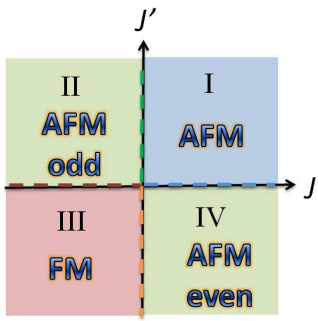}
\caption{ (Color online) Phase diagram for the spin-$1/2$ bond-alternating Ising chain. }
 \label{fig5}
\end{figure}

 We plot these order parameters as a function of the interaction parameter $\theta$ in Fig.~\ref{fig4}.
 It is shown clearly that each defined order parameter is non-zero
 for the region of each of the ordered phases, otherwise zero.
 This shows that the defined order parameters 
 characterize each of the ordered phases.
 Furthermore, the quantum phase transitions are seen to be first order
 by the sudden drop of the order parameters to zero at the
 discontinuous points of the fidelity revealed in Fig. \ref{fig2}.
 At these points the groundstates are infinitely degenerate.
 As a result, the spin-$1/2$ bond-alternating Ising model has
 the four ordered phases.
 As is shown in the phase diagram of Fig. \ref{fig5}, the system is clearly seen to be in the

 \begin{itemize}

 \item[(i)] AFM phase for $0 < \theta < \pi/2$ $(J > 0 \mbox{~and~} J' > 0)$,

 \item[(ii)] odd AFM phase for  $\pi/2 < \theta < \pi$ $(J < 0 \mbox{~and~} J' > 0)$,

 \item[(iii)] FM phase for $\pi < \theta < 3\pi/2$ $(J < 0 \mbox{~and~} J' < 0)$,

 \item[(iv)] even AFM phase for $3\pi/2 < \theta < 2\pi$ $(J > 0 \mbox{~and~} J' < 0)$.

 \end{itemize}
 The phase boundaries are the lines $J=0$ and $J'=0$.
 
 %%%%%%%%%%%%%%%%%%%%%%%%%%%fig 5%%%%%%%%%%%%%%%%%%%%%%%%%%%%%%%%%%%%
%%%%%%%%%%%%%%%%%%%%%%%%%%%fig 5%%%%%%%%%%%%%%%%%%%%%%%%%%%%%%%%%%%%

 Actually, if the system undergoes a full spontaneous symmetry breaking of the group
 $G = Z_2 \times Z_2$, there are four degenerate groundstates.
 However, we have detected only the two degenerate groundstates in
 each phase from the fidelity shown in Fig.~\ref{fig2} and the order parameters in Fig.~\ref{fig4}.
 As one can notice in Fig.~\ref{fig4},
 the odd and even AFM phases
 are clearly distinguished by the defined order parameters.
 These show that
 for the odd and even AFM phase,
 the system undergoes a partial spontaneous symmetry breaking of the
 group, which induces the two degenerate ground states for each phase.
 Depending on the sign of the interaction strengths $J$ and $J'$,
 the partial spontaneous symmetry breaking generates the different
 ordered phases, i.e., the odd and even AFM phases.
 Consequently, the odd and even AFM phases are distinguishable.

\section{Conclusion}

 We have investigated quantum fidelity in an infinite-size bond-alternating Ising chain
 by employing the iMPS representation with the iTEBD method.
By detecting the doubly degenerate groundstates for each phase
 by means of the quantum fidelity with an arbitrary
 reference state,
 it was shown that, for each phase, a different $Z_2$ symmetry is broken
 and the system state belongs to a different ordered phase.
 By also detecting the discontinuities of the quantum fidelity,
 we demonstrated that first-order quantum
 phase transitions occur between the ordered phases
 as the interaction parameter $\theta$ varies through $0 < \theta < 2\pi$.
 Based on the spin configurations from the characteristic properties of the spin correlations,
 the four defined local order parameters, including the even and the odd AFM order parameters,
 are shown to clearly characterize each phase and the
 existence of the first-order quantum phase transitions between the ordered phases.
 Consequently, by taking a simple and well known model as example, 
 we have demonstrated the usefulness of the quantum fidelity
 with an arbitrary reference state
 to investigate the nature of quantum phases
 without knowing a priori what type of internal order is present in a quantum many-body state.

\acknowledgments

 It is a pleasure to acknowledge Professor Huan-Qiang Zhou for encouragement and support.
 This work was supported by the National Natural Science Foundation
 of China (Grant No. 11374379).
 M.T.B. is supported by the 1000 Talents Program of China. His work is also partially supported by the Australian Research Council.

\end{document}